# Energy Efficiency in Multi-hop CDMA Networks: A Game Theoretic Analysis[*]

Sharon Betz and H. Vincent Poor
Department of Electrical Engineering
Princeton University
{sbetz,poor}@princeton.edu

## Abstract

*A game-theoretic analysis is used to study the effects of receiver choice on the energy efficiency of multi-hop networks in which the nodes communicate using Direct-Sequence Code Division Multiple Access (DS-CDMA). A Nash equilibrium of the game in which the network nodes can choose their receivers as well as their transmit powers to maximize the total number of bits they transmit per unit of energy is derived. The energy efficiencies resulting from the use of different linear multiuser receivers in this context are compared, looking at both the non-cooperative game and the Pareto optimal solution. For analytical ease, particular attention is paid to asymptotically large networks. Significant gains in energy efficiency are observed when multiuser receivers, particularly the linear minimum mean-square error (MMSE) receiver, are used instead of conventional matched filter receivers.*

## 1 Introduction

In a wireless multi-hop network, nodes communicate by passing messages for one another; permitting multi-hop communications, rather than requiring one-hop communications, can increase network capacity and allow for a more ad hoc (and thus scalable) system (with little or no centralized control). For these reasons, and because of their potential for commercial, military, and civil applications, wireless multi-hop networks have attracted considerable attention over the past few years. In these networks, energy efficient communication is important because the nodes are typically battery-powered and therefore energy-limited. Work on energy-efficient communication in these multi-hop networks has often focused on routing protocols; this work instead looks at power control and receiver design choices that can be implemented independently of (and thus in conjunction with) the routing protocol.

One approach that has been very successful in researching energy efficient communications in both cellular and multi-hop networks is the game-theoretic approach described in [1, 2]. Much of the game-theoretic research in multi-hop networks has focused on pricing schemes (e.g. [3, 4]). In this work, we avoid the need for such a pricing scheme by using instead a nodal utility function to capture the energy costs. It further differs from previous research by considering receiver design, as [5] does for cellular networks.

We propose a distributed noncooperative game in which the nodes can choose their transmit power and linear receiver design to maximize the number of bits that they can send per unit of power. After describing the network and internodal communications in Section 2, we derive the Nash equilibrium for this game, as well as for a set of games with set receivers, in Section 3. We then extend the asymptotic work of Tse and Hanly [6] to fit the multihop network structure in Section 4; we apply this in Section 5 to find the Pareto optimal solution in an asymptotically large, SINR-balanced network. Finally we present some numerical results and a conclusion in Sections 6 and 7.

## 2 System Model

Consider a wireless multi-hop network with $K$ nodes (users) and an established logical topology, where a sequence of connected link-nodes $l \in L(k)$ forms a route originating from a source $k$ (with $k \in L(k)$ by definition). Let $m(k)$ be the node after node $k$ in the route for node $k$. Assume that all routes that go through a node $k$ continue through $m(k)$ so that node $k$ transmits only to $m(k)$. Nodes communicate with each other using DS-CDMA with processing gain $N$ ($N$ chips per bit).

The signal received at a node $m$ (after chip-matched filtering) sampled at the chip rate over one symbol duration

---

[*]This research was supported in part by the U. S. Air Force Research Laboratory and in part by the Defense Advanced Research Projects Agency.

can be expressed as

$$\mathbf{r}^{(m)} = \sum_{k=1}^{K} \sqrt{p_k} h_k^{(m)} b_k \mathbf{s}_k + \mathbf{w}^{(m)} \qquad (1)$$

where $p_k$, $b_k$, and $\mathbf{s}_k$ are the transmit power, transmitted symbol, and (binary) spreading sequence for node $k$; $h_k^{(m)}$ is the channel gain between nodes $k$ and $m$; and $\mathbf{w}^{(m)}$ is the noise vector which is assumed to be Gaussian with mean $\mathbf{0}$ and covariance $\sigma^2 \mathbf{I}$. (We assume here $p_m = 0$.) Assume the spreading sequences are random, i.e., $\mathbf{s}_k = \frac{1}{\sqrt{N}}[v_1 \ldots v_N]^T$, where the $v_i$'s are independent and identically distributed (i.i.d.) random variables taking values $\{-1, +1\}$ with equal probabilities. Denote the cross-correlations between spreading sequences as

$$\rho_{kj} = \mathbf{s}_k^T \mathbf{s}_j, \qquad (2)$$

noting that $\rho_{kk} = 1$ for all $k$.

Let us represent the linear receiver at the $m$th node for the $k$th signature sequence by a coefficient vector $\mathbf{c}_k^{(m)}$. The output of this receiver can be written as

$$y = \mathbf{c}_k^T \mathbf{r}^{(m)} \qquad (3)$$
$$= \sqrt{p_k} h_k^{(m)} b_k \mathbf{c}_k^T \mathbf{s}_k + \sum_{j \neq k} \sqrt{p_j} h_j^{(m)} b_j \mathbf{c}_k^T \mathbf{s}_j + \mathbf{c}_k^T \mathbf{w}^{(m)}. \qquad (4)$$

The signal-to-interference-plus-noise ratio (SINR), $\gamma_k$, of the $k$th user at the output of receiver $m(k)$ is

$$\gamma_k = \frac{p_k h_k^{(m(k))^2} \left(\mathbf{c}_k^T \mathbf{s}_k\right)^2}{\sigma^2 \mathbf{c}_k^T \mathbf{c}_k + \sum_{j \neq k} p_j h_j^{(m(k))^2} \left(\mathbf{c}_k^T \mathbf{s}_j\right)^2}. \qquad (5)$$

Each user has a utility function that is the ratio of its effective throughput to its transmit power, i.e.,

$$u_k = \frac{T_k}{p_k}. \qquad (6)$$

Here, the throughput, $T_k$, is the net number of information bits sent by $k$ (generated by $k$ or any node whose route goes through $k$) and received without error at the intended destination, $m(k)$, per unit of time. (We assume that all the congestion control is done in the choice of routing.)

Following the discussion in [5], we will use

$$T_k = \frac{L}{M} R f(\gamma_k) \qquad (7)$$

where $L$ and $M$ are the number of information bits and the total number of bits in a packet, respectively (without loss of generality assumed here to be the same for all users); $R$ is the transmission rate, which is the ratio of the bandwidth to the processing gain and is taken for now to be equal for all users; and $f(\cdot)$ is an efficiency function that closely approximates the packet success rate. This efficiency function can be any increasing, continuously differentiable, sigmoidal[1] function with $f(0) = 0$ and $f(+\infty) = 1$. See [5] for more discussion of the efficiency function.

Using (7), (6) becomes

$$u_k = \frac{L}{M} R \frac{f(\gamma_k)}{p_k}. \qquad (8)$$

When the receiver used is a matched filter (MF) (i.e. $\mathbf{c}_k^{(m(k))} = \mathbf{s}_k$), the received SINR is

$$\gamma_k^{\text{MF}} = \frac{p_k h_k^{m(k)^2} \left(\mathbf{s}_k^T \mathbf{s}_k\right)^2}{\sigma^2 \mathbf{s}_k^T \mathbf{s}_k + \sum_{j \neq k} p_j h_j^{m(k)^2} \left(\mathbf{s}_k^T \mathbf{s}_j\right)^2} \qquad (9)$$

$$= \frac{p_k h_k^{m(k)^2}}{\sigma^2 + \sum_{j \neq k} p_j h_j^{m(k)^2} \rho_{kj}^2}. \qquad (10)$$

When the receiver is a linear minimum mean-squared error (MMSE) receiver, the filter coefficients and the received SINR are [7]

$$\mathbf{c}_k^{\text{MMSE}} = \frac{\sqrt{p_k} h_k^{m(k)}}{1 + p_k h_k^{m(k)^2} \mathbf{s}_k^T \mathbf{A}_k^{-1} \mathbf{s}_k} \mathbf{A}_k^{-1} \mathbf{s}_k \qquad (11)$$

and

$$\gamma_k^{\text{MMSE}} = p_k h_k^{m(k)^2} \mathbf{s}_k^T \mathbf{A}_k^{-1} \mathbf{s}_k, \qquad (12)$$

where

$$\mathbf{A}_k = \sigma^2 \mathbf{I} + \sum_{j \neq k} p_j h_j^{m(k)^2} \mathbf{s}_j \mathbf{s}_j^T. \qquad (13)$$

When the receiver is a decorrelator[2] (DE) (i.e. $\mathbf{C} = [\mathbf{c}_1 \cdots \mathbf{c}_K] = \mathbf{S}(\mathbf{S}^T \mathbf{S})^{-1}$ where $\mathbf{S} = [\mathbf{s}_1 \cdots \mathbf{s}_K]$), the received SINR is

$$\gamma_k^{\text{DE}} = \frac{p_k h_k^{m(k)^2}}{\sigma^2 \mathbf{c}_k^T \mathbf{c}_k}. \qquad (14)$$

For any linear receiver with all nodes' coefficients chosen independently of their transmit powers (including the MF and DE), as well as for the MMSE receiver,

$$\frac{\partial \gamma_k}{\partial p_k} = \frac{\gamma_k}{p_k}. \qquad (15)$$

---

[1] A continuous increasing function is sigmoidal if there is a point above which the function is concave and below which the function is convex.

[2] Here, we must assume that $K \leq N$.

## 3 The Noncooperative Power-Control Game

Let $\mathcal{G} = \left[\mathcal{K}, \{A_k\}, \{u_k\}\right]$ denote the noncooperative game where $\mathcal{K} = \{1, \ldots, K\}$ and $A_k = [0, P_{\max}] \times \mathbb{R}^N$ is the strategy set for the $k$th user. Here, $P_{\max}$ is the maximum allowed power for transmission. Each strategy in $A_k$ can be written as $\mathbf{a}_k = (p_k, \mathbf{c}_k)$, where $p_k$ and $\mathbf{c}_k$ are the transmit power and the receiver filter coefficients, respectively, of user $k$. Then the resulting noncooperative game can be expressed as the maximization problem for $k = 1, \ldots, K$:

$$\max_{\mathbf{a}_k} u_k = \frac{LR}{M} \max_{p_k, \mathbf{c}_k} \frac{f(\gamma_k(p_k, \mathbf{c}_k))}{p_k}, \quad (16)$$

where $\gamma_k$ is expressed explicitly as a function of $p_k$ and $\mathbf{c}_k$.

This is similar to the noncooperative power-control game in [5]; here, however, the channel gains are between pairs of nodes rather than between a node and the base-station.

Since the choice of receiver is independent of the transmit power and $f(\cdot)$ is an increasing function, the analysis of [5] applies, so the maximization from (16) becomes:

$$\max_{p_k, \mathbf{c}_k} \frac{f(\gamma_k(p_k, \mathbf{c}_k))}{p_k} = \max_{p_k} \frac{f(\max_{\mathbf{c}_k} \gamma_k(p_k, \mathbf{c}_k))}{p_k}. \quad (17)$$

Note that the MMSE receiver achieves the maximum SINR amongst all linear receivers, so that if a Nash equilibrium exists, at that equilibrium all receivers must be MMSE receivers. Then the maximization problem becomes

$$\max_{p_k} \frac{f(\gamma_k^{\text{MMSE}}(p_k))}{p_k}. \quad (18)$$

Let $\mathcal{G}_{\mathbf{C}} = \left[\mathcal{K}, \{[0, P_{\max}]\}, \{u_k\}\right]$ denote the noncooperative game that differs from $\mathcal{G}$ in that users cannot choose their linear receivers but are forced to use the receive filter coefficients $[\mathbf{c}_1 \cdots \mathbf{c}_K] = \mathbf{C}$ (which may be a function of the powers, $\mathbf{P}$). The resulting noncooperative game can be expressed as the following maximization problem for $k = 1, \ldots, K$:

$$\max_{\mathbf{a}_k} u_k = \max_{p_k} u_k(p_k, \mathbf{c}_k) = \frac{LR}{M} \max_{p_k} \frac{f(\gamma_k^{\mathbf{c}_k}(p_k))}{p_k} \quad (19)$$

where $\gamma_k^{\mathbf{c}_k}$ is expressed explicitly as a function of $p_k$. Then the maximization problem in (18) is one of the games $\mathcal{G}_{\mathbf{C}}$ when $\mathbf{C}$ is chosen to be the MMSE receivers.

For any $\mathbf{C}$ matrix (or $\mathbf{C}(\mathbf{P})$ for which (15) holds), the utility function for each user is maximized when

$$p_k = \min\{P_k, p_k^*\} \quad (20)$$

where $p_k^*$ is the unique positive number that satisfies

$$f(\gamma_k^{\mathbf{c}_k}(p_k^*)) = \gamma_k^{\mathbf{c}_k}(p_k^*) f'(\gamma_k^{\mathbf{c}_k}(p_k^*)). \quad (21)$$

As long as the users all have the same efficiency function,

$$\gamma_1^{\mathbf{c}_1}(p_1^*) = \ldots = \gamma_K^{\mathbf{c}_K}(p_K^*) = \gamma^* \quad (22)$$

where $\gamma^*$ is the unique positive number that satisfies

$$f(\gamma^*) = \gamma^* f'(\gamma^*). \quad (23)$$

Finally, since $\frac{f(\gamma_k)}{p_k}$ is quasi-concave[3] in $p_k$, we can use the result cited in [2, Appendix I]: $\mathcal{G}_{\mathbf{C}}$ has a Nash equilibrium and, as is the case in [5], it is unique. At this equilibrium, unless there is a node $k$ with $p_k^* > P_k$, the powers are such that the nodes are SINR-balanced (i.e. (22) holds).

Returning to the game $\mathcal{G}$, a similar result holds: there exists a unique equilibrium where all receivers are MMSE detectors and, if the power limit is high enough, the powers are SINR-balanced.

## 4 Asymptotically Large Systems: Extending the Tse-Hanly Equations to Multi-Hop Networks

Assume that the channel gains are independent. That is, in the asymptotic regime when $N, K \to \infty$ while $K/N = \beta$, the interferers' channel gains, $h_k^{(m)2}$ for all $m \neq k, m(k)$, are iid realizations of the random variable $\mathsf{G}$ with pdf $f_G$, and the primary channel gains, $h_k^{(m(k))2}$ for all $k$, are iid realizations of the random variable $\mathsf{H}$ with pdf $f_H$ (where $f_H(h) = 0 \forall h \leq 0$). Let $q = \mathbb{P}\{m(j) = m(k)\}$ for all $j \neq k$.

We can apply results from [6] to analyze the nodes' SINRs. Then we find a probability density function for $p$ such that in an asymptotically large system where all nodes have powers distributed by this function, with probability one all nodes have SINR of at least $\gamma$ for some $\gamma$. If this distribution is not unique, we choose the one that minimizes the nodes' powers. For simplicity, and since we are considering the asymptotic regime, we assume that the distribution of $p_k$ is independent of all channel gains except for $h_k^{(m(k))2}$. For convenience of notation, let $f_{p,H}(\cdot,\cdot) = f_{p_k, h_k^{(m(k))2}}(\cdot,\cdot)$, and note $\int_0^\infty f(p,h)dp = f_H(h)$ for all $h$. Then the joint density of $p_k$, $h_k^{(m(k))2}$, and $h_k^{(m(j))2}$ for $j \neq k$ is

$$f_{p_k, h_k^{(m(k))2}, h_k^{(m(j))2}}(p, h, g) = f_{p,H}(p, h)\delta(g - h)q \\ + f_{p,H}(p, h)f_G(g)(1 - q). \quad (24)$$

Applying the results from [6], when the receiver at node $k$ is a matched filter, decorrelator, or MMSE receiver, the random SINR at the receiver converges in probability as $N, K \to \infty$ while $K/N = \beta$. These asymptotic SINRs are uniquely described by the equations (where $j \neq k$):

$$\gamma^{\text{MF}} = \frac{p_k h_k^{(m(k))2}}{\sigma^2 + \beta \mathbb{E}\left[p_j h_j^{(m(k))2}\right]} \tag{25}$$

$$\gamma^{\text{DE}} = \begin{cases} \frac{p_k h_k^{(m(k))2}(1-\beta)}{\sigma^2}, & \alpha < 1; \\ 0, & \alpha \geq 1. \end{cases} \tag{26}$$

and

$$\gamma^{\text{MMSE}} = \frac{p_k h_k^{(m(k))2}}{\sigma^2 + \beta \int_0^\infty dp \int_0^\infty dg f_P(p) f_G(g) I(pg, p_k h_k^{(m(k))2}, \gamma^{\text{MMSE}})}, \tag{27}$$

where $I(a, b, c) = \frac{ab}{b+ac}$.

---

[3] A function is quasi-concave if there exists a point below which the function is nondecreasing and above which the function is non-increasing.

If the nodes choose their transmit powers so that the SINRs are balanced, the following theorem determines what SINRs are achievable at all receivers as well as the minimum transmit powers to achieve any achievable SINR when the nodes use the MMSE receiver, under the assumptions listed above.

**Theorem 4.1.** *A necessary and sufficient condition for an SINR, $\gamma$, to be achievable is for*

$$\beta \gamma q \frac{1}{1+\gamma} + \beta \gamma (1-q) \mathbb{E}\left[\frac{\mathsf{G}}{\mathsf{H} + \gamma \mathsf{G}}\right] < 1. \tag{28}$$

*When (28) holds, each user can achieve the desired SINR, $\gamma$, and the minimum power solution to do so is to assign each node, k, transmit power*

$$p_k = P_{\text{MMSE}}\left(h_k^{(m(k))2}, \gamma\right) \tag{29}$$

$$= \frac{1}{h_k^{(m(k))2}} \cdot \frac{\gamma \sigma^2}{1 - \beta \gamma q \frac{1}{1+\gamma} - \beta \gamma (1-q) \mathbb{E}\left[\frac{\mathsf{G}}{\mathsf{H}+\gamma \mathsf{G}}\right]}. \tag{30}$$

### 4.1 Proof of Theorem 4.1

We start with a lemma that is a straightforward consequence of the definition of $I(a, b, c)$.

**Lemma 4.2.** *For all positive real numbers $a_0, a, b, c$, $a_0 \leq a$ if and only if $I(a_0, b, c) \leq I(a, b, c)$.*

Then the proof follows.

*Proof.* To show necessity, assume that there is a pdf $f$ with $\int_0^\infty f(p, h) dp = f_H(h)$ for all $h$, such that in an asymptotically large system where all nodes have powers and primary channel gains distributed by $f$, with probability one all nodes have SINR when using an MMSE receiver of at least $\gamma$ for some set $\gamma$. Let $Q = \inf\{ph : f(p,h) > 0\}$. Then

$$\begin{aligned}\frac{Q}{\gamma} &\geq \sigma^2 + \beta \int_0^\infty dg \int_0^\infty dp \int_0^\infty dh f_{p_k h_k^{(m(k))2}, h_k^{(m(j))2}}(p, h, g) I(pg, Q, \gamma) \\ &= \sigma^2 + \beta q \int_0^\infty dp \int_0^\infty dh f_{p,H}(p,h) I(ph, Q, \gamma) \\ &+ \beta(1-q) \int_0^\infty dg \int_0^\infty dp \int_0^\infty dh f_{p,H}(p,h) f_G(g) I(ph\frac{g}{h}, Q, \gamma) \\ &\geq \sigma^2 + \beta q \int_0^\infty dp \int_0^\infty dh f_{p,H}(p,h) I(Q, Q, \gamma) \\ &+ \beta(1-q) \int_0^\infty dg \int_0^\infty dp \int_0^\infty dh f_{p,H}(p,h) f_G(g) I(Q\frac{g}{h}, Q, \gamma) \\ &= \sigma^2 + \beta q \frac{Q}{1+\gamma} + \beta(1-q) \int_0^\infty dg \int_0^\infty dh f_H(h) f_G(g) \frac{gQ}{h+\gamma g} \\ &= \sigma^2 + Q\beta q \frac{1}{1+\gamma} + Q\beta(1-q) \mathbb{E}\left[\frac{\mathsf{G}}{\mathsf{H}+\gamma \mathsf{G}}\right]. \end{aligned} \tag{31}$$

This implies that

$$Q\left(1 - \beta \gamma q \frac{1}{1+\gamma} - \beta \gamma (1-q) \mathbb{E}\left[\frac{\mathsf{G}}{\mathsf{H}+\gamma \mathsf{G}}\right]\right) \geq \gamma \sigma^2 > 0, \tag{32}$$

so $\beta \gamma q \frac{1}{1+\gamma} + \beta \gamma (1-q) \mathbb{E}\left[\frac{\mathsf{G}}{\mathsf{H}+\gamma \mathsf{G}}\right] < 1$, proving necessity.

When (28) holds, it is easy to show that $P_{\text{MMSE}}(h, \gamma)$ is positive for all primary channel gains, $h$. It is also straightforward to show that if each node, $k$, uses transmit power $P_{\text{MF}}\left(h_k^{(m(k))2}, \gamma\right)$, all nodes will achieve the SINR requirement, $\gamma$, finishing the proof of sufficiency.

Finally, consider any other joint distribution of powers and primary channel gains whose marginal distribution for $\mathsf{H}$ is $f_H$, and let $Q^*$ be the minimal received power in this distribution. Then by exactly the same argument as was used in the proof of necessity,

$$Q^* \geq \frac{\gamma \sigma^2}{1 - \beta \gamma q \frac{1}{1+\gamma} - \beta \gamma (1-q) \mathbb{E}\left[\frac{\mathsf{G}}{\mathsf{H}+\gamma \mathsf{G}}\right]} \tag{33}$$

$$= h P_{\text{MMSE}}(h, \gamma), \forall h > 0. \tag{34}$$

This means that assigning powers according to $P_{\text{MMSE}}$ does indeed give the minimal power solution. □

## 5 A Global Optimization Problem

A useful global optimization problem is

$$\max \sum_{k=1}^K \alpha_k u_k = \frac{L}{M} R \max \sum_{k=1}^K \frac{\alpha_k f(\gamma_k)}{p_k}, \tag{35}$$

where the $\alpha_k$'s are set weighting variables. This problem is equivalent to finding a Pareto-optimal solution of the game.

According to [5], even in the special case of a cellular system where $L(k) = \{k\}$ for all nodes $k = 1, 2, \ldots, K$ and all nodes are transmitting to the base-station, "Pareto-optimal solutions are, in general, difficult to obtain." For simplicity, we restrict the problem by requiring that the solution is "fair": all nodes have equal receiver output SINRs (i.e. SINR-balancing), so $\gamma = \gamma_1 = \gamma_2 = \ldots \gamma_K$.

With this assumption, (35) becomes

$$\frac{L}{M} R \max f(\gamma) \sum_{k=1}^{K} \frac{\alpha_k}{p_k}. \tag{36}$$

For the matched filter, we can apply (5) with $m = m(k)$ to see that the users' SINRs are equal if and only if

$$\left(B + \left(\frac{1}{\gamma} + 1\right) D\right) \mathbf{p}(\gamma) = \sigma^2 \mathbf{1} \tag{37}$$

where $B$ is a $K$ by $K$ matrix with entries $B_{kj} = -h_j^{(m(k))^2} \rho_{kj}^2$, $D$ is $K$ by $K$ diagonal matrix with diagonal entries $D_{kk} = h_k^{(m(k))^2}$, and $\mathbf{1}$ is a vector of $K$ ones.

The SINR that maximizes (36) is the $\gamma$ that satisfies

$$0 = \frac{\partial}{\partial \gamma} \left[ f(\gamma) \sum_{k=1}^{K} \frac{\alpha_k}{p_k(\gamma)} \right] \tag{38}$$

$$= \frac{\partial}{\partial \gamma} [f(\gamma)] \sum_{k=1}^{K} \frac{\alpha_k}{p_k(\gamma)} - f(\gamma) \sum_{k=1}^{K} \frac{\alpha_k}{p_k^2(\gamma)} \frac{\partial}{\partial \gamma} [p_k(\gamma)], \tag{39}$$

where $p_k(\gamma)$ and $\frac{\partial}{\partial \gamma} [p_k(\gamma)]$ are the $k$th elements of

$$\mathbf{p}(\gamma) = \sigma^2 \left(B + \left(\frac{1}{\gamma} + 1\right) D\right)^{-1} \mathbf{1} \tag{40}$$

and

$$\frac{\partial}{\partial \gamma} [\mathbf{p}(\gamma)] = \sigma^2 (\gamma B + (1+\gamma) D)^{-1} D (\gamma B + (1+\gamma) D)^{-1} \mathbf{1}. \tag{41}$$

For the decorrelator, it is easy to show that the noncooperative results are equal to the globally optimal results, since the users' achieved SINRs are independent of all the powers of all interferers.

Finally, for the MMSE receiver, we can apply the results from Section 4. In a large system, if all users choose their transmit powers based on the values of $h_k^{(m(j))}$ for $m(j) \neq m(k)$ only through the average of these interference gains and if we use the assumptions of Section 4, the SINR is approximated by

$$\gamma_k^{\text{MMSE}} \simeq \frac{p_k h_k^{(m(k))^2}}{\sigma^2 + \frac{1}{N} \sum_{j \neq k} I(p_j h_j^{(m(k))^2}, p_k h_k^{(m(k))^2}, \gamma_k^{\text{MMSE}})}. \tag{42}$$

Any $\gamma_k$ which satisfies $\frac{\partial \gamma_k}{\partial p_k} = \frac{\gamma_k}{p_k}$ is a solution to (42).

Then the power for user $k$ to achieve the SINR $\gamma^*$ is

$$p_k^{\text{MMSE}} = \frac{1}{h_k^{(m(k))^2}} \frac{\gamma^* \sigma^2}{1 - \beta \gamma^* \left(q \frac{1}{1+\gamma^*} + (1-q)\zeta(\gamma^*)\right)}, \tag{43}$$

where $\zeta(\gamma)$ is the mean value of $\frac{G}{H + \gamma G}$ in the network. Equal received SINRs amongst the users is achieved with minimum power consumption when $p_k h_k^{(m(k))^2} = \kappa(\gamma)$ is constant for all $k$ and

$$\kappa(\gamma) = \frac{\gamma \sigma^2}{1 - \beta \gamma \left(q \frac{1}{1+\gamma} + (1-q)\zeta(\gamma)\right)}. \tag{44}$$

Then, (36) can be expressed as

$$\frac{L}{M} R \left( \sum_{k=1}^{K} \alpha_k h_k^{(m(k))^2} \right) \max_{\gamma} \frac{f(\gamma)}{\kappa(\gamma)}. \tag{45}$$

The solution to $\max_\gamma \frac{f(\gamma)}{\kappa(\gamma)}$ must satisfy $\frac{\partial}{\partial \gamma} \left( \frac{f(\gamma)}{\kappa(\gamma)} \right) = 0$. Combining this with (44) gives the equation that must be satisfied by the solution to the maximization problem in (45):

$$f(\gamma) = \gamma f'(\gamma) \left( 1 - \frac{\frac{\beta q \gamma}{(1+\gamma)^2} + \beta(1-q)\gamma\zeta(\gamma)}{1 - \frac{\beta q \gamma^2}{(1+\gamma)^2} - \beta(1-q)\gamma\zeta(\gamma)} \right). \tag{46}$$

If $\zeta(\gamma) \ll 1$, then the equation is approximately the same as in the cellular case [5] with $K/N \to \beta q$. Then, the ability to use multiple hops to communicate, and therefore reduce transmit power, has similar results to reducing the system load; furthermore, for a large range of values of $\beta q$, the MMSE target SINRs for the noncooperative game and for the Pareto-optimal solution are close.

# 6 Numerical Results

Consider a multi-hop network with $K = 100$ nodes distributed randomly in a square 500 meters by 500 meters surrounding an access point in the center. We use a simple routing scheme where all nodes transmit to the closest node that is closer to the access point (or the access point if that is closest). We assume that each packet contains 100 bits of data and no overhead ($L = M = 100$); the transmission rate is $R = 100$ kb/s; the thermal noise power is $\sigma^2 = 5 \times 10^{-16}$ Watts; the channel gains are distributed with a Rayleigh distribution with mean $0.3 d^{-2}$, where $d$ is the distance between the transmitter and receiver; and the processing gain is $N$. We use the same efficiency function as [5], namely $f(\gamma) = (1 - e^{-\gamma})^M$.

Table 1 shows the average utility for four representative sets of randomly chosen spreading sequences, one for each of $N$ = 50, 100, 200, and 300, comparing the mean utility under the various power choice method discussed above. Table 2 shows the target SINRs for the socially optimal results displayed in Table 1.

|  | MF | DE | MMSE |
|---|---|---|---|
| $N = 50$ |  |  |  |
| non-coop. | 0 |  | $1.198 \times 10^{10}$ |
| soc. opt. | $2.025 \times 10^{-14}$ |  | $1.199 \times 10^{10}$ |
| $N = 100$ |  |  |  |
| non-coop. | 0 | $1.095 \times 10^{8}$ | $1.417 \times 10^{10}$ |
| soc. opt. | $1.050 \times 10^{-4}$ | $1.095 \times 10^{8}$ | $1.417 \times 10^{10}$ |
| $N = 200$ |  |  |  |
| non-coop. | 0 | $7.459 \times 10^{9}$ | $1.476 \times 10^{10}$ |
| soc. opt. | $2.512 \times 10^{-10}$ | $7.459 \times 10^{9}$ | $1.476 \times 10^{10}$ |
| $N = 300$ |  |  |  |
| non-coop. | 0.2056 | $1.001 \times 10^{10}$ | $1.493 \times 10^{10}$ |
| soc. opt. | $1.351 \times 10^{9}$ | $1.001 \times 10^{10}$ | $1.493 \times 10^{10}$ |

**Table 1. Mean utilities for four representative sets of spreading sequences.**

| $N$ | MF | DE | MMSE |
|---|---|---|---|
| 50 | 0.87 |  | 6.39 |
| 100 | 1.31 | 6.47 | 6.43 |
| 200 | 0.99 | 6.47 | 6.45 |
| 300 | 5.03 | 6.47 | 6.46 |

**Table 2. Socially optimal SINRs for the same four representative sets of spreading sequences.**

The socially optimally implemented MF receiver performs poorly in heavily-loaded systems, while the non-cooperative implementation fails to achieve non-zero utility except in the case with the lightest load. Even in the case where $\beta = 1/3$, the mean utility for the socially optimal MF receiver is less than a tenth of the MMSE receiver's mean utility. Using the DE receiver (for which we require that $K \geq N$), as was noted in Section 5, there is no difference between the non-cooperative and socially optimal results: both cases have the same target SINR and thus the same mean utility. For the MMSE receiver, this difference between the mean utility in the non-cooperative and socially optimal implementations is very small. Finally, the DE and MMSE receivers both significantly outperform the MF receiver in all four of these cases. There is, however, a price to pay in using the better-performing receivers: these receivers require more information at every node as well as significantly more computation. These issues will be addressed further in later research.

# 7  Conclusion

We have analyzed the cross-layer issue of energy-efficient communication in multi-hop networks using a game theoretic method. Focusing on linear receivers, we have derived the transmit power levels that results in a Nash equilibrium for multiple receiver designs, showing that at this equilibrium the users are SINR-balanced. We then generalized the important asymptotic work of Tse and Hanly to allow for the case where users and their interferers may be transmitting to different locations, keeping the cellular example as a special case. We applied these asymptotic results, as well as exact results for the MF and DE receivers, to find the equations for the SINR-balanced Pareto-optimal solution. We showed that the MMSE receiver is the optimal receiver and that in many cases the non-cooperative MMSE receiver results are quite close to the socially optimal results.